\documentclass[twocolumn,showpacs,preprintnumbers,amsmath,amssymb,superscriptaddress]{revtex4-1}
\usepackage{graphicx}

\begin{document}

\title{Tunnel magnetoresistance in alumina, magnesia and composite tunnel barrier magnetic tunnel junctions}

\author{Oliver Schebaum}
\author{Volker Drewello}
\author{Alexander Auge}
\author{G\"unter Reiss}
\affiliation{Thin films and physics of nanostructures, Bielefeld University, Germany}
\author{Markus M\"unzenberg}
\author{Henning Schuhmann}
\author{Michael Seibt}
\affiliation{I. and IV. Physikalisches Institut, Georg-August-Universit\"at G\"ottingen, Germany}
\author{Andy Thomas}
	\email{andy.thomas@uni-bielefeld.de}
\affiliation{Thin films and physics of nanostructures, Bielefeld University, Germany}


\begin{abstract}
Using magnetron sputtering, we have prepared Co-Fe-B/tunnel barrier/Co-Fe-B magnetic tunnel junctions with tunnel barriers consisting of alumina, magnesia, and magnesia-alumina bilayer systems. The highest tunnel magnetoresistance ratios we found were 73\% for alumina and 323\% for magnesia-based tunnel junctions. Additionally, tunnel junctions with a unified layer stack were prepared for the three different barriers. In these systems, the tunnel magnetoresistance ratios at optimum annealing temperatures were found to be 65\% for alumina, 173\% for magnesia, and 78\% for the composite tunnel barriers. The similar tunnel magnetoresistance ratios of the tunnel junctions containing alumina provide evidence that coherent tunneling is suppressed by the alumina layer in the composite tunnel barrier.
\end{abstract}

\maketitle


\section{Introduction}
In recent years, magnetic tunnel junctions (MTJs) have garnered much interest due to the large number of possible applications, such as magnetic random access memory (MRAM) and magnetic logic \cite{Wolf:2001p95,Reiss:2006p6536,Chappert:2007p108}. In most cases, a large tunnel magnetoresistance (TMR) ratio is desired.

The TMR effect was discovered at room temperature in alumina-based magnetic tunnel junctions \cite{Moodera:1995p66,Miyazaki:1995p5737}, as this material was well studied from tunneling experiments with superconductors \cite{Giaever:1960p74, Meservey:1970p77}. The highest measured TMR ratio has gradually increased over time and has been measured to be as high as 80\% at room temperature \cite{Wang:2004p6583,Wei:2007p60}.

In addition to alumina, other materials, such as strontium-titanate  \cite{Oguz:2009p6151,Thomas:2005p3421} and titanium-oxide \cite{Bibes:2003p6461}, were used as tunnel barriers in MTJs.  In 2001, higher TMR ratios were predicted for Fe/MgO/Fe systems with crystalline tunnel barriers and electrodes  \cite{Butler:2001p4537, Mathon:2001p6616} and were subsequently experimentally verified \cite{Parkin:2004p71,Yuasa:2004p106}. Now, TMR ratios of up to 604\% are observed in MgO-based MTJs at room temperature \cite{Ikeda:2008p3496}.

In this manuscript, we investigate magnetic tunnel junctions with alumina and magnesia barriers and compare them to MTJs with alumina-magnesia bilayers as the tunnel barrier. For all the junctions studied, we examine the transport properties as a function of the annealing temperature.

The goal of our investigation is to find evidence for non-coherent tunneling processes in the bilayer magnetic tunnel junctions. We expect to find TMR ratios of the bilayer that are comparable to the pure alumina system, since the coherence is destroyed by the alumina layer. This is in contrast to simple spin-polarization models by e.g.\ Julli\`{e}re that would predict values in-between the values for MgO and alumina junctions \cite{Julliere:1975p11}.
\section{Preparation}
We studied MTJs with tunnel barriers that consist either of a single layer of Al$_2$O$_3$ or MgO. Additionally, we investigated MgO - Al$_2$O$_3$ bilayer structures as tunnel barrier materials. The thickness of each layer forming a tunnel barrier was always larger than 1.2\,nm, to avoid pinholes.
The layer stack and the annealing process varied for the respective samples and are provided in the results and discussion sections.
The samples were structured using UV optical lithography and Ar-ion beam etching, with element sizes between 25\,$\mu$m$^2$ and 700\,$\mu$m$^2$.
The measurements were performed using a standard two terminal setup. A constant voltage of 10\,mV was applied during the resistance vs. magnetic field measurements.
\section{Magnesia and alumina reference samples}
\begin{figure}
    \includegraphics[width=\linewidth]{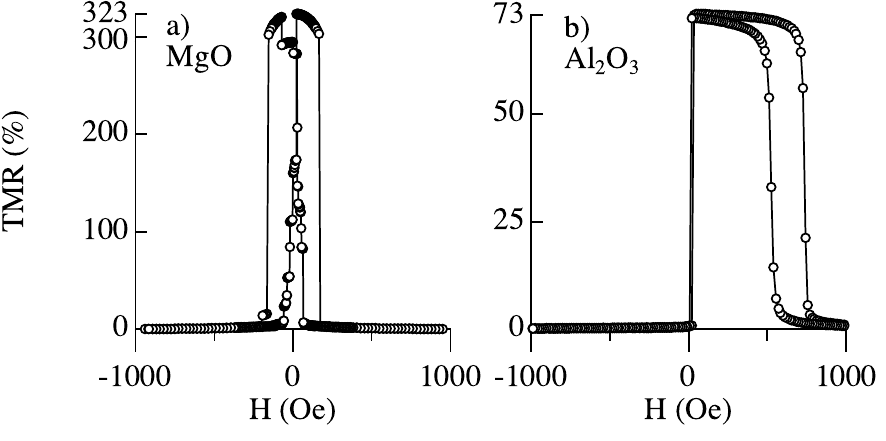}
    \caption{Tunnel magnetoresistance vs.\ magnetic field curves of MTJs with Co-Fe-B electrodes for different tunnel barrier materials. In part (a), a major loop of MgO based MTJs is presented, and in part (b), a major loop of Al$_2$O$_3$ based MTJs is shown. The highest observed TMR ratios are 323\% for MgO-based MTJs and 73\% for Al$_2$O$_3$-based MTJs.}
    \label{best_ones}
\end{figure}

We prepared two reference samples to optimize the alumina and MgO preparation processes. The layer stacks, sputter conditions, and annealing temperatures were adjusted to yield the highest TMR values. The TMR vs. magnetic field ($H$) curves of these samples are shown in Figure~\ref{best_ones}.
First, we discuss the details of the MgO preparation, after which the alumina sample preparation will be outlined.

The layer stack of the MgO sample was Ta 20/Co-Fe-B 5.3/MgO 2.4/Co-Fe-B 3.2/Ta 20 (all values in nm) with a spin valve structure, i.e., hard-soft switching of the electrodes. The sample was deposited by magnetron sputtering in a sputter system with a base pressure of $1\times 10^{-9}$\,mbar. The crystallization of the barrier and the electrode-barrier interfaces was initiated by a post-annealing step in a vacuum furnace with an operating pressure of $2 \times 10^{-7}$\,mbar. With an annealing temperature of 450$^\circ$C for one hour, the highest TMR ratios of about 320\% were achieved. The magnetoresistance vs. magnetic field curve of one optimized MgO-based junction is depicted in Figure~\ref{best_ones}(a).

In Figure~\ref{best_ones}(b), the TMR vs. H loop of the optimized Al$_2$O$_3$-based MTJ is shown. The layer stack of this sample was Ta 5/Cu 30/Ta 5/Cu 5/Mn-Ir 12/Co-Fe-B 4/Al 1.2 + oxidation/Co-Fe-B 4/Ni-Fe 3/Ta 5/Cu 20/Au 50 (all numbers in nm). The samples were fabricated using DC and RF magnetron sputtering in an automatic sputtering system with a base pressure of $1 \times 10 ^{-7}$\,mbar. The metallic Al layer was sputtered and post-oxidized by remote plasma oxidation in a separate oxidation chamber. The details of the alumina preparation are presented in Ref. \cite{Thomas:2003p6630}. The exchange coupling of the hard magnetic electrode was activated in a post-annealing and field-cooling step with an in-plane magnetic field of 6500\,Oe and in the same furnace used for the MgO samples. The optimal annealing temperature was 275$^\circ$C for five minutes. Here, we obtained a TMR ratio of 73\%.
Next, we have to combine the preparation processes of the samples to be able to better compare all of the sample types: the alumina-based ones, MgO-based ones, and the bilayer systems.
\section{Unified layer stack}
The unified layer stacks consisted of Ta/Ru/Ta/Ru/Mn$_{17}$Ir$_{83}$ under-layers, followed by a Co$_{40}$Fe$_{40}$B$_{20}$(2.5 nm)/tunnel barrier/Co$_{40}$Fe$_{40}$B$_{20}$(3 nm) tri-layer. A Ta/Ru/Au cover stack provided protection for the upper electrode and a reliable electrical contact.

To form a hard magnetic electrode, the lower Co-Fe-B layer was exchange coupled to the underlying anti-ferromagnetic Mn-Ir layer. The exchange bias was activated in a post-annealing and field-cooling step similar to the alumina samples. The samples were annealed for one hour to initialize the crystallization of the MgO layers.

The annealing temperatures for the MgO samples were lower than before, therefore yielding lower TMR ratios. This is due to the necessity of having the preparation process for the MgO samples as similar as possible to the aluminum process. In the unified stack, the TMR decrease at higher temperatures is caused by manganese diffusion out of the Mn-Ir layer towards the barrier \cite{Hayakawa:2006p72} and overcompensates the crystallization of the barrier and barrier/electrode interfaces.

\begin{figure}
	\includegraphics[width=\linewidth]{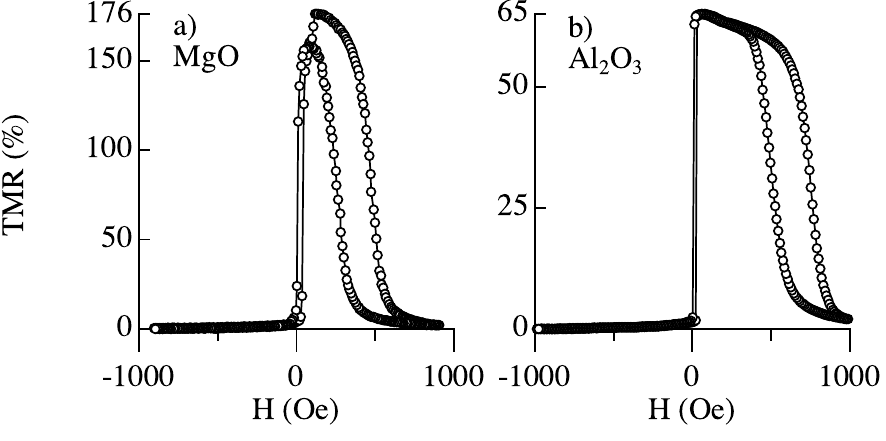}
	\caption{In part (a), a major loop of MgO based MTJs is depicted, and in part (b), a major loop of Al$_2$O$_3$ based MTJs is shown. The junctions of the unified layer stack exhibit maximum TMR ratios of 176\% and 65\% for MgO and Al$_2$O$_3$, respectively.}
	\label{fig:unified_mtjs}
\end{figure}

The annealing temperatures were chosen to produce the highest TMR ratios and were measured to be 325$^\circ$C for the MgO samples and 275$^\circ$C for the alumina-based samples. The major loops of the unified MgO and alumina-based MTJs are shown in Figure~\ref{fig:unified_mtjs}. The TMR ratio of the two similarly prepared stacks are 176\% and 65\% for the MgO and alumina-based samples, respectively.

Although the TMR ratio of the MgO-based sample decreased, it is still larger by a factor of 2.5 than the ratio for the alumina-based junctions. The TMR ratio measurements provide evidence for symmetry filtering \cite{Butler:2001p4537, Mathon:2001p6616} due to either coherent or non-coherent tunneling. The amplitude of the TMR ratio enables us to easily distinguish the two cases.

The single insulating layer of the reference MTJs was replaced by a MgO/Al$_2$O$_3$ bilayer system to form the composite barrier. First, a 1.4\,nm-thick MgO layer was directly deposited by RF sputtering from an MgO sputter target. Then, the second layer was formed from a 1.2\,nm-thick post-oxidized Al film.

\begin{table}[h]
    \centering
    \begin{tabular}{lccc}
    \hline
    $T_{\text{A}}$\,($^\circ$C for one hour) & $275$ & $300$ & $350$ \\
    \hline
    TMR (\%) & $57.7$ & $77.8$ & $35.5$\\
    \hline
    \end{tabular}
    \caption{Dependence of the TMR ratio of the composite-tunnel barrier-based MTJs on the annealing temperature $T_{\text{A}}$.}
    \label{tab:TMR_vs_Ta}
\end{table} 

\begin{figure}
	\includegraphics[width=\linewidth]{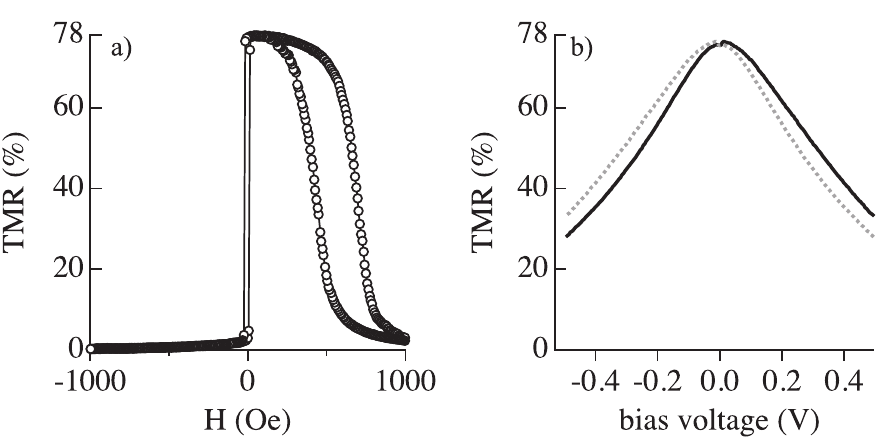}
	\caption{a) TMR major loop for the MgO/Al$_2$O$_3$ composite tunnel barrier junction. The highest TMR ratio of 78\% is attained for annealing temperatures of 300$^\circ$C. b) Voltage dependence of the TMR-ratio (solid line). The grey dots indicate the mirror image to indicate the asymmetry of the curve.}
	\label{composite}
\end{figure}

\begin{figure}
    \includegraphics[width=\linewidth]{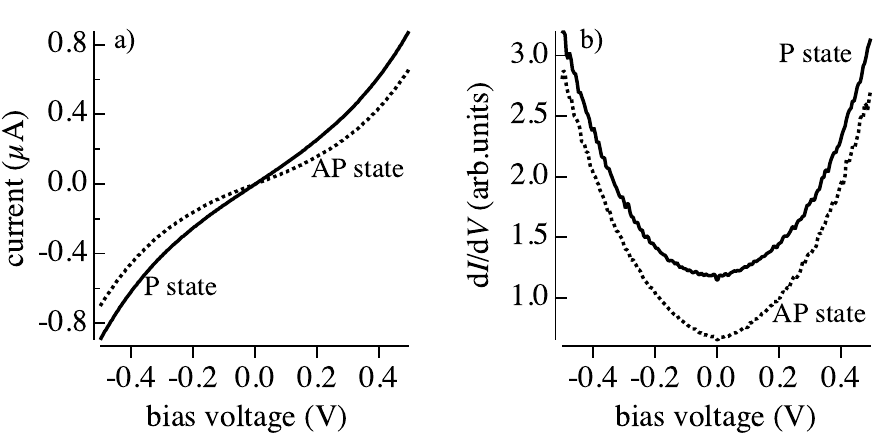}
    \caption{a) $I$ vs $V$ and b) $\text{d}I/\text{d}V$ vs $V$ characteristics in parallel and antiparallel state of the composite barrier tunnel junction.}
    \label{figure4}
\end{figure}

In Table~\ref{tab:TMR_vs_Ta}, we show the dependence of the TMR ratio on the annealing temperature for the MTJs with the MgO/Al$_2$O$_3$ composite tunnel barrier. The composite barrier MTJs show a maximum TMR ratio of 78\% at an optimum annealing temperature of 300$^\circ$C. The corresponding major loop of the optimized junction is depicted in Figure~\ref{composite} (a). In Figure~\ref{composite} (b) the bias voltage dependence of the TMR ratio is shown. The grey dots indicate the mirrored curve to point out the asymmetry. The TMR$-V$ data are calculated from the $I-V$ data in parallel and antiparallel state depicted in Figure~\ref{figure4}~(a). In Figure~\ref{figure4}~(b) the d$I$/d$V-V$ curves of the parallel and antiparallel states are shown.

In the curves shown in Figure~\ref{figure4} no particular variations can be seen. The general characteristics are very similar to MgO- and alumina-based juntions. In some cases, subtle features can be deduced out of IET-spectra of MTJs \cite{Drewello:2009p6787}. However, no IET-spectra could be taken in the composite MTJs due to the high resistance of the thick tunnel barrier.

It is crucial for our investigation to obtain not only two barriers on top of each other, but also each without pinholes. Therefore, the bilayer tunnel barrier is thicker than the tunnel barriers of the single alumina or magnesia-based MTJs. Nonetheless, Figure~\ref{best_ones}(a) proves, in principle, that coherent tunneling is possible in thicker tunnel barriers.
\section{Discussion}
The highest TMR ratio with bilayer tunnel barriers in the present work is 78\%. This ratio is in the range of values reported for Al$_2$O$_3$ based MTJs. This is true for the low temperature values as well. A TMR ratio of 118\% was measured at 20\,K, compared to 114\% for the pure alumina MTJs \cite{Schmalhorst:2007p6659}. The small increase in the TMR ratio might be attributed to the higher interface quality of the Co-Fe-B/MgO layer. The TMR ratio is still smaller than the highest value observed in alumina junctions \cite{Wang:2004p6583,Wei:2007p60}.

This is strong evidence that the effect of symmetry filtering is eliminated. Any magnetic tunnel junction with an amorphous alumina layer destroys the coherent tunneling process. This is consistent with our measurements and the observations of Sukegawa et al.\ \cite{Sukegawa:2010p6866}. In MTJs with a crystalline Fe/spinel MgAl$_2$O$_4$/Fe structure, TMR ratios of 117\% at room temperature have been reported, which exceeds the highest reported ratios for alumina-based MTJs. This can again be explained by the symmetry filtering due to the crystalline tunnel barrier.

The TMR ratio of the composite barrier MTJ drops from 78\% at low voltages to 28\% for -0.5\,V and 33\% at +0.5\,V as shown in Figure~\ref{composite}~(b). Such an asymmetry can neither be observed in pure alumina \cite{Wang:2004p6583} nor in MgO based MTJs. In pure MgO-based MTJs the TMR ratio drops from 320\% at low voltages to 150\% at $\pm0.5$\,V (not shown). For MgO, alumina, and composite tunnel barrier MTJs the TMR drops to roughly half of its value at a bias voltage of 0.5\,V.

Another simple explanation for the TMR ratio in MTJs was given by Julli\`{e}re in Ref.~\cite{Julliere:1975p11}. In Julli\`{e}re's model, only the effective spin polarizations of different ferromagnetic/insulator combinations contribute to the TMR ratio. Assuming this model, one would expect a TMR ratio of about 100\% for an MTJ with one Co-Fe-B/MgO and one Al$_2$O$_3$/Co-Fe-B interface. This is in contradiction to the observed results. 

There are only a small number of other reports on MgO/Al$_2$O$_3$ composite tunnel barriers. Theoretical and experimental studies have demonstrated high barrier asymmetries for such systems. A TMR ratio of 7\% and considerable asymmetry in the current-voltage characteristics have been reported for MTJs with Co electrodes \cite{deButtet:2006p5832}. The asymmetry in our MTJs that is indicated in Figure~\ref{composite}~(b) is not as pronounced. The large discrepancy in the TMR ratios (a factor of ten) suggests that extrinsic differences in the sample preparation (e.g., deposition techniques in this particular case) are responsible for the different results.
\section{Summary}
In summary, we have investigated the transport properties of MTJs with tunnel barriers consisting of single layers and bilayers of Al$_2$O$_3$ and MgO. The highest observed TMR ratio (78\%) of the bilayer at room temperature is on the order of the highest reported values for MTJs with Al$_2$O$_3$ tunnel barriers in other works. This indicates that the observed limitation of the TMR ratio in Al$_2$O$_3$-based MTJs is caused by incoherent tunneling through the amorphous Al$_2$O$_3$ layer.
\section{Acknowledgments}
The authors would like to thank C.\ A.\ Jenkins from Mainz University for helpful discussions. We also would like to acknowledge the MIWF of the NRW state government and the German Research Foundation DFG for financial support.
\end{document}